\documentclass[11pt, letterpaper]{article}
\usepackage[utf8]{inputenc}
\usepackage[margin=1.5cm]{geometry}
\usepackage{titlesec}
\usepackage{tabu}
\usepackage{enumitem}
\usepackage{threeparttable}
\usepackage{textgreek}
\usepackage{multirow}
\usepackage{amssymb}
\usepackage{xcolor}
\usepackage{color,soul}
\usepackage{siunitx}

\newlist{selectlist}{itemize}{2}
\setlist[selectlist]{label=$\square$,leftmargin=*,noitemsep,topsep=0pt}

\usepackage{lmodern}

\usepackage[main=english]{babel}	
\usepackage{url} 
\usepackage{eurosym} 
\usepackage{graphicx} 
\usepackage{epstopdf} 
\usepackage{caption}
\usepackage{verbatim} 
\usepackage{indentfirst} 

\usepackage{hyperref}
\hypersetup{
    colorlinks=true,
    linkcolor=blue,
    filecolor=magenta,      
    urlcolor=blue,
    citecolor=blue,
    bookmarksnumbered=true,
    bookmarksopen=true,
    bookmarksopenlevel=1,
    pdftitle={Four-channel radio-frequency signal generator programmed by an open-source Arduino-based control system via single or quad Serial Peripheral Interface},
    pdfstartview={FitH} 
}
 
\titleformat{\section}[block]{\hspace{1em}\bfseries}{\thesection.}{0.5em}{} 
\titleformat{\subsection}[block]{\hspace{1em}}{\thesubsection}{0.5em}{}

\begin{document}
\begin{flushleft}

\setlength{\parindent}{0pt}
\setlength{\parskip}{10pt}

\textbf{Four-channel radio-frequency signal generator programmed by an open-source Arduino-based control system via single or quad Serial Peripheral Interface}

\vskip 0.2cm

\textbf{Michele Sorelli\textsuperscript{a,b}, Marco Marchetti\textsuperscript{c}, Pietro Ricci\textsuperscript{a,b,d}, Domenico Alfieri\textsuperscript{c}, Vladislav Gavryusev\textsuperscript{a,b,*}, Francesco Saverio Pavone\textsuperscript{a,b,e}}\\
\vskip 0.2cm

\textbf{a University of Florence, Department of Physics and Astronomy, Sesto Fiorentino, 50019, Italy\\
b European Laboratory for Non-Linear Spectroscopy, Sesto Fiorentino, 50019, Italy\\
c L4T-Light4Tech, Sesto Fiorentino, 50019, Italy\\
d University of Barcelona, Department of Applied Physics, Barcelona, 08028, Spain\\
e National Institute of Optics, Sesto Fiorentino, 50019, Italy}\\ 
\vskip 0.1cm

\textbf{vladislav.gavryusev@unifi.it ‘twitter: \url{@VGavryusev}’}\\
\vskip 0.2cm


\textbf{Abstract}

Radio-frequency (RF) signal generators are standard laboratory equipment and a wide-range of open-source and commercial devices exists to address their many applications. Nonetheless, only few expensive and proprietary solutions can be re-configured within a wide frequency band and triggered on a micro-second timescale. Such specifications are required for applications that use variable radio-frequencies to generate programmed mixed signals, to control processes or states and to precisely steer laser beams using acousto-optical devices, tasks often needed in industrial manufacturing, atomic and molecular physics or microscopy.

Here we present an open-source low-cost Arduino-based control system that can store up to millions of commands received from a computer and then perform reliable high-speed programming of an arbitrary device under its control (DUC) via a single- or quad-wire Serial Peripheral Interface. The software architecture operates as a real-time state machine, making it easily extensible and adaptable to any DUC. Each configuration change can be triggered either externally or internally, reaching $\approx\SI{1}{\mega\hertz}$ rates when using a Teensy 4.1 Arduino-compatible board. Leveraging this flexible system, we developed a programmable four-channel RF signal generator, based on an Analog Devices 9959 Evaluation board, and we demonstrated its capability and validated its performance.
\vskip 0.2cm



\newpage
\textbf{Specifications table}\\
\vskip 0.2cm
\tabulinesep=1ex
\begin{tabu} to \linewidth {|X|X[3,l]|}
\hline  \textbf{Hardware name} & 
  \textit{Programmable four-channel radio-frequency signal generator}\\
  \hline \textbf{Subject area} & 
  \begin{itemize}[noitemsep, topsep=0pt]
  \item \textit{Engineering and material science}
  \item \textit{Physics}
  \item \textit{Open source alternatives to existing infrastructure}
  \end{itemize}
  \\
  \hline \textbf{Hardware type} & 
  \vskip 0.1cm  
  \begin{itemize}[noitemsep, topsep=0pt]
  \item \textit{Electrical engineering and computer science}
  \item \textit{External device control and programming}
  \end{itemize}
  \\ 
\hline \textbf{Closest commercial analog} &
  \textit{Multi-channel programmable radio-frequency signal generators, such as \href{https://www.wieserlabs.com/products/radio-frequency-generators/WL-FlexDDS-NG}{\underline{Wieserlabs WL-FlexDDS-NG}}.}\\
\hline \textbf{Open source license} &
  {CC-BY Attribution-ShareAlike 4.0 International (CC BY-SA 4.0)}\\
\hline \textbf{Cost of hardware} &
  \textit{641.79 \euro}\\
\hline \textbf{Source file repository} & 
  \textit{\href{https://doi.org/10.17632/hvwyz5yhh2.1}{\underline{Mendeley Data https://doi.org/10.17632/hvwyz5yhh2.1}}}\\
\\\hline
\end{tabu}
\end{flushleft}

\newpage
\section{Hardware in context}\label{sec:intro}
Radio-frequency (RF) signal generators are devices designed to produce continuous and pulsed signals in the RF and microwave domains with defined and adjustable frequency, phase and amplitude. Often they provide one or several methods to modulate these properties in order to create a continuous-wave (CW) output, single pulses, pulse trains or more complex waveforms. These devices have a wide-range of applications, both in industrial, commercial and laboratory settings, spanning from wireless communication, automated test equipment, imaging and spectroscopy for healthcare to experiments in neuroscience \cite{Juutilainen2011,Meneghetti2020,Yaghmazadeh2022}, biophysics \cite{DuemaniReddy2008,Corsetti2021}, microscopy \cite{Saggau1998,Duocastella2020}, particle and nuclear physics \cite{Demarteau2016}, atomic and molecular physics \cite{Pruttivarasin2015,Donnellan2019,Bertoldi2020}, and quantum simulation and computing \cite{Arute2019,Altman2021,Amico2021}. There is no single general purpose instrument that is able to cover such breadth of scopes and technical requirements, instead many open-source and commercial solutions have been developed to address one or several tasks, with their specific strengths, limitations and cost.

Recently, RF equipment has found vast application in industrial manufacturing, atomic physics and microscopy to precisely steer laser beams using RF-driven acousto-optical devices \cite{Duocastella2020,Gavryusev2019,Ricci2020,Ricci2022} and to control processes or states of matter that are sensitive to this frequency domain \cite{Arute2019}, either directly through electromagnetic radiation emitted by an antenna or indirectly by driving electronic or opto-electronic equipment. Achieving a fine degree of control often requires to generate complex sequences of single RF tones, multi-frequency waveforms and rapid frequency sweeps that interleave on microsecond timescales and span a wide band of several hundreds $\si{\mega\hertz}$. Most often a single RF output is not sufficient and typically four channels or more have to be used concurrently, while respecting stringent phase-coherence and sub-$\si{\micro\second}$ synchronization conditions. 

Arbitrary waveform generators (AWG) are a very flexible class of RF devices that can fulfill these demands. Commercial solutions \cite{Signatek,SpectrumInstrumentation,Allcock2021} tend to have a substantial Cost Per Channel (CPC) in the range of 2000-3000 \euro, while laboratory developed solutions \cite{Baig2013,Bowler2013,Govorkov2014} are more affordable.  These projects are based on field-programmable gate arrays (FPGA) that are very flexible and adaptable to changing requirements, but often their frequency bandwidth is limited to few tens of $\si{\mega\hertz}$. Besides, they present usability constraints because their software interfaces are often not trivial to program and integrate into existing laboratory control systems.

RF devices that employ Direct Digital Synthesis (DDS) to produce high spectral purity single-tones and frequency sweeps are a second valid class of solutions. Current state-of-the-art technology allows to manufacture DDS chips that can synthesize $\SI{1.4}{\giga\hertz}$ sine-waves, such as the Analog Devices (AD) AD9914 \cite{AD9914,AD9914PCBZ}, or even reach $\SI{4.2}{\giga\hertz}$ \cite{Zhang2014}, but their CPC is rather high in the range of 800-1000 \euro. More affordable generators are built using one or several DDS chips that cost approximately 50 \euro\space each and can provide a $\SIrange{200}{400}{\mega\hertz}$ output, such as the four channel AD9959 \cite{AD9959,AD9959PCBZ} and the single channel AD9910 \cite{AD9910,AD9910PCBZ} chips. 
Many commercial products have been developed using these or similar DDS technologies and the most accessible multi-channel equipment \cite{GraAndAfch,Novatech,ModularSystemControls,AAOptoElectronic} has a CPC within 50-200 \euro, but is severely limited in the Output Reprogramming Rate ($ORR \leq \SI{10}{\kilo\hertz}$). Higher speed RF drivers reach an $ORR\geq \SI{1}{\mega\hertz}$ by leveraging a built-in memory that can store thousands of consecutive settings received from a computer, allowing output stabilization and amplification, and complex modulation schemes, with the downside of a significantly increased CPC ranging from 500 \euro  \cite{MoglabsQRF,SinaraUrukul,Kasprowicz2022} to 2000-3000 \euro  \cite{SpinCoreDDS300,SpinCoreDDS1000,Wieserlabs,WieserlabsDual,MoglabsXRF}. 
Several research groups have developed in-house DDS-based signal generators to reach an affordable CPC of 250-400 \euro, while preserving specifications comparable to the best commercial products or even adding customized functionality, such as digital input-output channels or a complete experiment timing and control system \cite{Pruttivarasin2015,Perego2018,Prevedelli2019,Donnellan2019,Bertoldi2020,Allcock2021}. Like for AWGs, FPGAs have been used to program and control the DDS chips with $ORR\geq \SI{1}{\mega\hertz}$, sub-$\si{\micro\second}$ jitter triggering and providing a large command memory. Some designs have been presented without disclosing the implementation details \cite{Liang2009,Li2016,Perego2018,Donnellan2019}, while others have released all material as open-source \cite{Pruttivarasin2015,Bertoldi2020,Kasprowicz2022}. Whilst complying with the license terms, the latter choice enables any user to reproduce, enhance and adapt the equipment to their own specific requirements way beyond what a proprietary solution can allow, while substantial economic savings may be obtained \cite{Pearce2020}. Furthermore, the original developers receive recognition for their work and may benefit from improvements contributed by the wider community of users. 

Recently, the development of microcontroller units integrated into a development board (MCU) with ARM central processing units (CPU) running at clock rates of tens of $\si{\mega\hertz}$ has paved the way to an alternative to FPGAs for achieving negligible jitter, substantial command storage capacity and $ORR\geq \SI{1}{\mega\hertz}$. This approach presents several advantages: MCUs such as the broad Arduino-compatible family of devices \cite{Arduino} are programmed in C++, instead of requiring knowledge of the more specialized VHDL and VERILOG languages, which lowers the usage barrier and vastly broadens the potential user community. Next, MCUs have longer product lifetimes than FPGAs and provide notably greater software and hardware compatibility between different product generations, including application programming interfaces, communication buses and even dimensional blueprints which even enable drop-in replacements or upgrades. This feature stimulates the creation and continued support of extensive software libraries that ease the firmware development and of interoperable off-the-shelf hardware plugins and extensions. Additionally, MCUs provide several standardized communication interfaces that may allow interacting with arbitrary devices under control (DUC) that present the same protocol, such as the Inter-Integrated Circuit (I\textsuperscript{2}C) and Serial Peripheral Interface (SPI). Interestingly, this may obviate the need of realizing custom printed circuit boards (PCB) to pilot a DDS or another DUC. All together these aspects help to reduce development time and CPC expenditure.

This paper presents an open-source low-cost Arduino-based control system that can store millions of commands received from a computer (PC) via Universal Serial Bus (USB) and perform reliable high-speed programming of an arbitrary device under its control through a single- or quad-wire SPI. We use a Teensy 4.1 development board \cite{Teensy41} as the MCU. The Teensy 4.1 is an  Arduino-compatible development board with an ARM Cortex-M7 CPU running at $\SI{600}{\mega\hertz}$, which grants it several times more computing power than provided by the $\SI{84}{\mega\hertz}$ ARM Cortex-M3 CPU of an Arduino Due \cite{ArduinoDue}. The controller software architecture operates as a real-time state machine, making it easily extensible and adaptable to any DUC. Each configuration change can be triggered either externally or internally, with rates up to $\approx\SI{200}{\kilo\hertz}$ when using the standard SPI library. Outstandingly, if the standard single-wire SPI library is replaced by a custom implementation that benefits from port masking optimizations, the $ORR$ rate doubles in single-wire SPI mode and reaches $\approx\SI{1}{\mega\hertz}$ with quad-wire SPI communication, surpassing all prior MCU-based solutions \cite{ModularSystemControls,GraAndAfch}. Leveraging this flexible system, we developed a low-cost programmable four-channel RF signal generator, based on an Analog Devices 9959 evaluation board \cite{AD9959,AD9959PCBZ}, and demonstrated its capability, validating its performance for all use-cases that require $ORR\leq \SI{1}{\mega\hertz}$, low jitter and stand-alone operation with large command memory. The total system cost is currently 641.79 \euro, resulting in a notably low CPC of 160.45 \euro.

First, we present an overview of the hardware and software that composes our system. Then, all design files and materials are provided and discussed, along with build and operation instructions. Finally, we measure and validate the performance of our open-source programmable four-channel RF signal generator controlled by an Arduino-based MCU.

\section{Hardware description}\label{sec:hw_descr}
\subsection{Overview}\label{sec:hw_descr:overview}

\begin{figure}[!ht]
	\centering
	\includegraphics[width=1\textwidth]{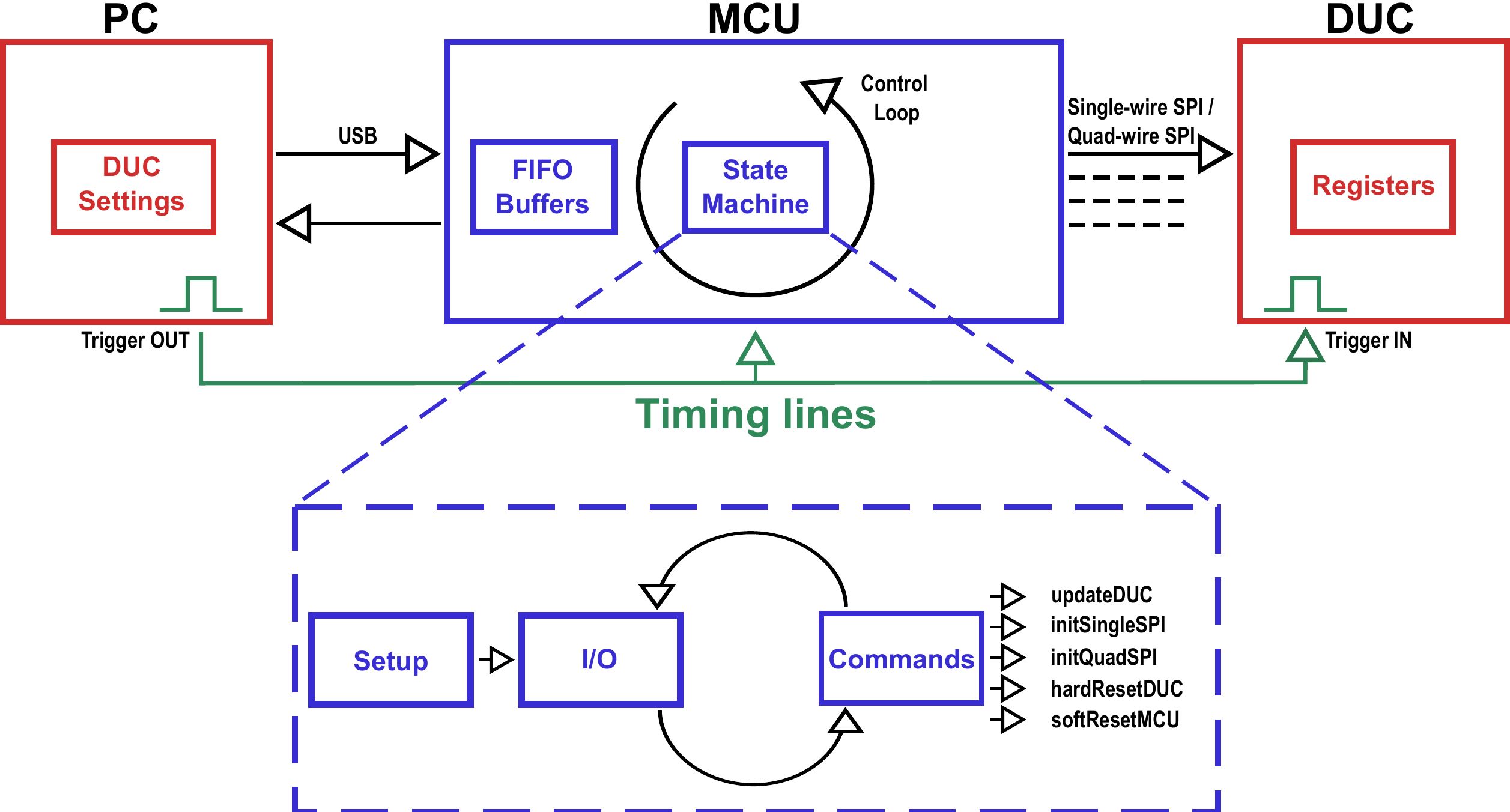}
	\caption{Block diagram of the MCU-based system, designed to pilot an arbitrary device-under-control via single-wire or quad-wire SPI, while receiving commands from a computer via USB. The MCU is programmed to run a real-time state machine that reacts to external events and provides a fast command memory for the DUC.}
	\label{fig:scheme}
\end{figure}

The design of the programmable four-channel RF signal generator that we developed is based on a general purpose and flexible architecture made of four main elements: the user determines on a PC the set of commands and settings that the MCU has to program into the DUC with precise timing supplied either internally or externally through a timing system. The MCU software implements a real-time state machine that can receive new settings and commands during run-time, store them in a fast internal memory, control the DUC and react to external events. The block diagram of the system is shown in Fig.~\ref{fig:scheme} and its elements will be presented in detail in the following.

\subsection{Hardware components}\label{sec:hw_descr:hw}

The central hardware element of our architecture is the MCU hosted on a development board because it interfaces the PC with the DUC and acts as a hardware and software abstraction layer, lifting from the user the need of knowing the technical details of the device to be controlled. Thanks to its built-in memory, the MCU can work both in tandem with the PC, receiving new commands during run-time, or standalone after storing the sequence of commands and settings to be applied to the DUC. Furthermore, it can be either externally triggered or perform timing functions itself since almost all microcontroller development boards provide many digital and some analog input and output (IO) channels.

We chose to use a Teensy 4.1 development board \cite{Teensy41} as the MCU because it is currently the most powerful Arduino-compatible solution \cite{Arduino}. Its software is programmed in C++ (described in detail in the following Sec.~\ref{sec:hw_descr:soft_mcu}), a language which many developers and researchers can work with, granting a large potential user community, instead of the more specialized VHDL and VERILOG languages that are required to operate most FPGAs. The Teensy has an ARM Cortex-M7 CPU with a float point math unit running at $\SI{600}{\mega\hertz}$ which provides several times more computing power than the $\SI{84}{\mega\hertz}$ ARM Cortex-M3 CPU of an Arduino Due \cite{ArduinoDue}. This enables to process input data and IO communications with sub-$\si{\micro\second}$ timescale latency and jitter, both via USB 2.0 (with the PC) and SPI interfaces (with the DUC) or I\textsuperscript{2}C. It has a total of 55 IO pins with different characteristics, making it capable of reacting to external events and providing precise triggering to the DUC. Several sets of four IO pins can be toggled together very quickly using a hardware-based port mapping optimization, a feature that we leveraged to implement a custom single- and quad-wire SPI that significantly increased the data rate with the DUC, as demonstrated in Sec.~\ref{sec:val}. Furthermore, this MCU has a large memory subsystem consisting of a 1024K random-access-memory (RAM), 7936K Flash, 4K EEPROM and with the option of QSPI memory expansion up to 16~MByte by soldering two extra RAM chips. Notably, the interaction with the PC can be realized not only via USB, but also through the Ethernet 10/100 Mbit interface. This option has not been implemented in our project, but should be easy to add through an existing library. 

The computer has no special hardware requirements and can run any operating system that supports the Python Jupyter notebook software package and USB or Ethernet communication. The timing system has to provide five interrupts for the MCU and five triggers for the DUC (as shown in Fig.~\ref{fig:pinconn} and presented in detail in Sec.~\ref{sec:build_instr}), totaling a requirement of ten digital transistor-transistor logic (TTL) channels, and can be either a standalone device or a board integrated in the PC, such as the National Instruments (NI) board PCI-6251 \cite{NationalInstrumentsPCI6251} that we used in our implementation. This PCI Multifunction I/O Device has 16 analog inputs (16-Bit, 1.25 MegaSamples/s), 2 analog outputs and 24 digital IO, out of which 8 can be timed up to 10 MHz. One of its TTL lines is logically combined through an OR gate circuit (Fig.~\ref{fig:schematicaux}(B)) with a TTL line from the MCU to generate a logic signal that triggers the output update on the DUC.

The MCU can be engineered and programmed to control any DUC that provides a supported interface, such SPI and  I\textsuperscript{2}C. Since we aimed to realize an open-source low-cost programmable four-channel RF signal generator, we leveraged the capabilities of the Arduino-based control system to drive an AD9959/PCBZ evaluation board \cite{AD9959PCBZ}. This off-the-shelf equipment uses an AD9959 DDS chip \cite{AD9959} which provides four synchronous RF outputs that can reach $\SI{200}{\mega\hertz}$, generate single tones and linear frequency/phase/amplitude sweeps, and has independent frequency/phase/amplitude control. This RF source presents a narrow output spectrum with low phase-noise and it has $\SI{12}{\hertz}$ or better frequency tuning resolution, 14-bit phase offset resolution and 10-bit output amplitude scaling resolution. To operate, it requires two supply voltages ($+\SI{1.8}{\volt}$ DDS core and  $+\SI{3.3}{\volt}$ serial I/O) and a stable and spectrally narrow sinusoidal frequency reference that can be provided either by soldering on-board a crystal oscillator or supplying a clock signal externally. The clock signal is fed into a phase-locked loop (PLL) where a selectable $4\times$ to $20\times$ REF\_CLK multiplier is applied to generate the internal clock signal that provides timing to all internal components of the DDS chip and determines the maximum RF frequency that can be generated without incurring in spectrum distortions, which is 40\% of this rate. As clock signal source, we used an external $\SI{25}{\mega\hertz}$ temperature compensated crystal oscillator with a frequency stability of $\pm 280$ parts per billion, assembled following the schematic presented in Fig.~\ref{fig:schematicaux}(A).

The MCU is powered directly by the PC through the USB cable, while all other components of our programmable four-channel RF signal generator receive the required $+\SI{3.3}{\volt}$ and $+\SI{1.8}{\volt}$ supply voltages from the  power supply circuit that consists of an AC/DC and a DC/DC converters, as depicted in Fig.~\ref{fig:schematicaux}(C).

\subsection{MCU software}\label{sec:hw_descr:soft_mcu}
The MCU software loaded into the Teensy 4.1 board implements a real-time state machine able to listen to any incoming serial USB communication from PC, and to execute a set of predefined commands in response to external interrupt events. Here we will present its architecture in more detail.

The \textit{setup()} function initializes the state machine when the software is first loaded into the MCU board and configures the digital pins needed to interact with the DUC and the external timing system. Moreover, upon setup a hard reset signal is issued to the DUC to force its internal registers to their default state. Next, depending on the selected reference clocking configuration, the internal PLL-based clock multiplier factor is optionally set by programming the Function Register 1 (FR1) of the DUC via SPI. Finally, the serial USB communication between the PC and the MCU board is initialized, and external interrupt requests (IRQs) are enabled.

The I/O block handling the reception of subsequent DUC settings from the PC is realized by two functions that are executed consecutively within the \textit{loop()} function of the MCU sketch. Respectively, they read and parse incoming data strings, and push the received DUC configurations into the first in, first out (FIFO) buffers allocated for each of the four DDS channels. Similarly, there are separate sets of FIFO buffers designated for each specific channel register to be programmed into the DUC (refer to the AD9959 datasheet for a complete description of these registers).
The size of the FIFO buffers was set to 4500 elements to maximize the usage of the dynamic memory normally accessible on the Teensy 4.1 board (RAM1). This size corresponds to the maximum number of output configurations that can be activated on the DUC without the need for a new USB communication when all four RF channels are simultaneously updated and operated in frequency sweep mode, which represents the most memory-consuming scenario. If such use-case would not be foreseen, the number of elements could be increased by adapting the software to the expected workload. Furthermore, this threshold may be heightened by exploiting the secondary RAM space (RAM2) through dynamic memory allocation via the \textit{malloc()} function and by QSPI memory expansion, as mentioned in the previous subsection.
Before writing data to these memory buffers, the received floating point frequency values (namely single-tone frequencies, frequency sweep limit frequencies and step sizes) are converted to the tuning data words to be programmed into the respective internal registers of the DUC. In fact, performing this conversion directly when receiving the data strings from the PC enables a faster refresh of the DUC, increasing the maximum achievable $ORR$.

External interrupt events are used to react asynchronously to specific user commands set on the PC and they are disabled during the USB and SPI communication sessions. They are activated on the rising edge of the toggling of the designated interrupt pins that are connected to the external timing lines, as presented in detail in Sec.~\ref{sec:build_instr}. The interrupt service routines (ISRs) are the following:
\begin{itemize}
\item \textit{initSingleSPI()}: enables the custom single-wire SPI communication mode (default), with the simultaneous control of the designated SDIO\_0 and SCLK pins implemented via direct digital port manipulation;
\item \textit{initQuadSPI()}: enables the custom quad-wire SPI communication mode, which uses an efficient hardware implementation based on the fast simultaneous control of the designated SDIO pins via direct digital port manipulation;
\item \textit{softResetMCU()}: resets all MCU state variables to their default values, and clears the content of the FIFO buffers that store DDS tuning words previously received from the PC and not yet activated on the DUC;
\item \textit{hardResetDUC()}: issues a master reset pulse on the active-high reset pin of the DUC, reinitializing its internal registers to their default state. Afterwards, all channels are set to the single-tone mode of operation with their default 0x00 frequency tuning words, i.e. $\SI{0}{\mega\hertz}$. Then, if required, the routine programs the PLL multiplier factor back to the user-set value;
\item \textit{updateDUC()}: this routine first retrieves and interprets a channel configuration byte header associated with each data string communicated to the MCU, properly adjusting the mode of operation (i.e., single-tone or linear frequency sweep) of the relevant channels which need to be updated. More specifically, the least significant nibble of this header points out the channels operating in frequency sweep mode, whereas the most significant one identifies the reprogrammed channels, in accordance with the structure of the DUC \textit{channel selection register}, for a more efficient bit manipulation.
Next, the ISR reads the new single-tone or sweep tuning words from the FIFO buffers of the channels which need to be updated, and transfers them to the DUC via SPI. 
In order to activate the received tuning words and effectively change the signal generator output, a trigger must be issued on the I/O update pin of the DUC after the SPI communication is complete.
We implemented two options to generate this TTL: either the NI provides it, which requires careful synchronization with the execution of the update ISR, or the MCU itself supplies a pulse at the end of the data transfer. This latter approach, which we termed ``auto update" mode, requires to activate a variable in the source code to be engaged, as described in item 7 of Sect.~\ref{sec:op_instr}. Both options can work alternatively without hardware modification by using the digital logic OR gate circuit.
\end {itemize}

\subsection{Computer software}\label{sec:hw_descr:soft_pc}
The Jupyter notebook included in the software repository provides an easy-to-use user interface, written in Python 3.8, which allows to communicate to the MCU board the frequency settings to be consecutively activated on the DUC. All the functions included in the notebook are exhaustively documented, with docstrings complying with the PEP 257 convention. 

As detailed in Section~\ref{sec:op_instr}, users have only to edit the configuration lists within the “Input to AD9959 DDS” cell at the top of the notebook to customize the desired sequence of DUC settings. Whereas the specified single-tone frequency values can be directly transferred to the board, the generation of linear frequency sweeps by the AD9959/PCBZ evaluation board requires a conversion step of the user input. The desired positive or negative frequency slopes have to be translated into discrete time steps and intermediate frequency step sizes that are respectively applied when sweeping up or down the output frequency. This is performed by a dedicated block of chirp configuration functions which minimize the residual between the input chirp parameter and the slope of the linear sweep actually produced by the encoded sweep ramp rate and delta-tuning words, by iteratively selecting the adopted time step among a predefined list of increasing programmable values related to the DUC internal clock rate. Furthermore, also the transient phase leading to the maximum start frequency in the case of a falling frequency sweep, or back to the minimum frequency of a rising frequency sweep, must be similarly programmed. In the present implementation, this was accomplished so that the whole sweeping range was covered in the minimum time step allowed by the DUC, i.e. $\SI{8}{\nano\second}$ at the peak $\SI{500}{\mega\hertz}$ clock rate, thus producing a quasi-instantaneous recovery of the user defined starting value.

The developed Jupyter notebook finally features a set of functions devoted to handling the USB communication with the MCU. These take the single-tone and frequency sweep data related to the user-defined consecutive configurations of the DUC output channels, generate the channel mode byte header described in Sect.~\ref{sec:hw_descr:soft_mcu}, and composes the data strings which are finally encoded and transferred via USB to the MCU.

\subsection{Device usefulness potential}\label{sec:hw_descr:usefullness}
\noindent
In summary, the presented open-source programmable RF signal generator provides the following benefits:
\begin{itemize}
\item[$\bullet$]It has four outputs that can reach $\SI{200}{\mega\hertz}$ and operate in single-tone or frequency sweep modes.
\item[$\bullet$]The generator output can be reprogrammed very quickly, up to $ORR\approx\SI{1}{\mega\hertz}$.
\item[$\bullet$]The hardware design and software are completely open-source, allowing easy extension and customization.
\item[$\bullet$]All hardware components are off-the-shelf, for a total system cost of 641.79 \euro~and a 160.45 \euro~ CPC, very competitive with commercial and lab-built RF generators with similar specifications and applications.
\item[$\bullet$]The MCU software realizes a full control system that is adaptable to drive many other devices.
\end{itemize}

\section{Design files summary}\label{sec:design}
\noindent
The design files are stored in the \href{https://dx.doi.org/10.17632/hvwyz5yhh2.1}{Mendeley repository} and grouped in three folders:
\begin{itemize}
\item[$\bullet$]{Software: it contains the MCU sketch and the PC to MCU communication code.}
\item[$\bullet$]{Electronic Schematic: it contains the files describing the electronic design of the system.}
\item[$\bullet$]{Documentation: it contains the datasheets for the components of the system.}
\end{itemize}
\noindent
The key design files necessary to build and operate the system are the following:
\vskip 0.2cm
\tabulinesep=1ex
\noindent
\begin{tabu} to \linewidth {|X[1.35,1]|X[0.7,1]|X[0.6,1]|X[1.55,1]|} 
\hline
\textbf{Design filename} & \textbf{File type} & \textbf{Open source license} & \textbf{Location of the file} \\\hline
complete\_code\_pc\_mcu.zip & code archive from GitHub \cite{Github} & MIT & \href{https://dx.doi.org/10.17632/hvwyz5yhh2.1}{Mendeley}/Software/ \\\hline
python\_pc.ipynb & Jupyter notebook & MIT & \href{https://dx.doi.org/10.17632/hvwyz5yhh2.1}{Mendeley}/Software/python\_pc/ \\\hline
teensy\_mcu.ino & Arduino sketch & MIT & \href{https://dx.doi.org/10.17632/hvwyz5yhh2.1}{Mendeley}/Software/teensy\_mcu/ \\\hline
CircularBuffer-1.3.3 & Arduino library & GNU GPL v3 & \href{https://dx.doi.org/10.17632/hvwyz5yhh2.1}{Mendeley}/Software/teensy\_mcu/lib/ \\\hline
BOM.ods & OpenDocument spreadsheet & CC BY-SA 4.0 & \href{https://dx.doi.org/10.17632/hvwyz5yhh2.1}{Mendeley}/Electronic Schematic/ \\\hline
Kicad\_Circuits.zip & Kicad project & CC BY-SA 4.0 & \href{https://dx.doi.org/10.17632/hvwyz5yhh2.1}{Mendeley}/Electronic Schematic/ \\\hline
ElectronicSchematicsAll.pdf & PDF & CC BY-SA 4.0 & \href{https://dx.doi.org/10.17632/hvwyz5yhh2.1}{Mendeley}/Electronic Schematic/ \\\hline
CircuitsClockORgatePWR.pdf & PDF & CC BY-SA 4.0 & \href{https://dx.doi.org/10.17632/hvwyz5yhh2.1}{Mendeley}/Electronic Schematic/ \\\hline
PinConnections.pdf & PDF & CC BY-SA 4.0 & \href{https://dx.doi.org/10.17632/hvwyz5yhh2.1}{Mendeley}/Electronic Schematic/ \\\hline
\end{tabu}


\section{Bill of materials summary}\label{sec:bom}
\vskip 0.2cm
\tabulinesep=1ex
\noindent
\begin{tabu} to \linewidth {|X[2.0cm]|X[5cm]|X[1.5cm]|X[1.4cm]|X[1.4cm]|X[1.6cm]|X[2.5cm]|}
\hline
\textbf{Designator} & \textbf{Component} & \textbf{Number} & \textbf{Unit cost (\euro)} & \textbf{Total cost (\euro)} & \textbf{Source of materials} & \textbf{Material type} \\\hline

MCU & Teensy 4.1 development board & 1 & 31.53 & 31.53 & \href{https://www.pjrc.com/store/teensy41.html}{PJRC} & Other \\\hline

DUC & AD9959/PCBZ & 1 & 494.80 & 494.80 & \href{https://www.analog.com/en/design-center/evaluation-hardware-and-software/evaluation-boards-kits/eval-ad9959.html\#eb-buy}{Analog Devices} & Other \\\hline

Clock & CTS 535L250X2GT5 TCXO $\SI{25}{\mega\hertz}$ CLP SNW & 1 & 19.69 & 19.69 & \href{https://www.digikey.it/en/products/detail/cts-frequency-controls/535L250X2GT5/10711662}{Digi-Key} & Other \\\hline

OR & SN74LVC1G3208DBVR AND/OR Logic gate & 1 & 0.38 & 0.38 & \href{https://www.digikey.it/en/products/detail/texas-instruments/SN74LVC1G3208DBVR/863609}{Digi-Key} & Other \\\hline

PWR1 & LRS-50-3.3 AC/DC Converter $\SI{3.3}{\volt}$ $\SI{33}{\watt}$ & 1 & 15.92 & 15.92 & \href{https://www.digikey.it/en/products/detail/mean-well-usa-inc/LRS-50-3-3/7705049}{Digi-Key} & Other \\\hline

PWR2 & TPS82671EVM-646 DC/DC Converter $\SI{1.8}{\volt}$ & 1 & 23.96 & 23.96 & \href{https://www.digikey.it/en/products/detail/texas-instruments/TPS82671EVM-646/2441410}{Digi-Key} & Other \\\hline

ProtBoard & SBBTH1506-1 prototype board & 3 & 1.09 & 3.27 & \href{https://www.digikey.it/en/products/detail/chip-quik-inc/SBBTH1506-1/5978222}{Digi-Key} & Other \\\hline

J1, J2, J3, J4 & RF2-04A-T-00-50-G SMA female jack through hole & 4 & 1.84 & 7.36 & \href{https://www.digikey.it/en/products/detail/adam-tech/RF2-04A-T-00-50-G/9830588}{Digi-Key} & Other \\\hline

CBsma & Cable SMA-SMA male RG-316 $\SI{0.5}{\meter}$ & 2 & 14.60 & 29.20 & \href{https://www.digikey.it/en/products/detail/cinch-connectivity-solutions-johnson/415-0029-MM500/6579658}{Digi-Key} & Other \\\hline

CBusb & Cable USB 2.0 A male to micro B male $\SI{5}{\meter}$ & 1 & 7.99 & 7.99 & \href{https://www.digikey.it/en/products/detail/assmann-wsw-components/AK67421-5/2175143}{Digi-Key} & Other \\\hline

C1 & Capacitor $\SI{10}{\nano\farad}$ $\SI{50}{\volt}$ & 1 & 0.19 & 0.19 & \href{https://www.digikey.it/en/products/detail/kemet/C410C103K5R5TA7200/818228}{Digi-Key} & Ceramic \\\hline

C2, C4 & Capacitor $\SI{100}{\nano\farad}$ $\SI{50}{\volt}$ & 2 & 0.22 & 0.44 & \href{https://www.digikey.it/en/products/detail/kemet/C322C104K5R5TA7303/12701413}{Digi-Key} & Ceramic \\\hline

C3 & Capacitor $\SI{1}{\nano\farad}$ $\SI{50}{\volt}$ & 1 & 0.26 & 0.26 & \href{https://www.digikey.it/en/products/detail/tdk-corporation/FG18C0G1H102JNT06/5802785}{Digi-Key} & Ceramic \\\hline

C5 & Capacitor $\SI{47}{\micro\farad}$ $\SI{25}{\volt}$ & 1 & 0.22 & 0.22 & \href{https://www.digikey.it/en/products/detail/kemet/ESK476M035AC3EA/9448273}{Digi-Key} & Electrolytic \\\hline

TVS & SA7.0A-E3/54 Zener diode & 1 & 0.48 & 0.48 & \href{https://www.digikey.it/en/products/detail/vishay-general-semiconductor-diodes-division/SA7-0A-E3-54/2146115}{Digi-Key} & Semiconductor \\\hline

JP1, JP2 & SPC02SYAN jumper & 2 & 0.10 & 0.20 & \href{https://www.digikey.it/en/products/detail/sullins-connector-solutions/SPC02SYAN/76375}{Digi-Key} & Other \\\hline

CNm & Header Connector 32 position 2.54mm through hole & 2 & 0.79 & 1.58 & \href{https://www.digikey.it/en/products/detail/sullins-connector-solutions/PRPC032SAAN-RC/2775222}{Digi-Key} & Other \\\hline

CNf & Receptacle Connector 32 position 2.54mm through hole & 2 & 2.16 & 4.32 & \href{https://www.digikey.it/en/products/detail/mill-max-manufacturing-corp/310-47-132-41-001000/7364043}{Digi-Key} & Other \\\hline

\end{tabu}\\
\vskip 0.2cm
\noindent
The total cost of the components is 641.79 \euro, which leads to a CPC of 160.45 \euro.

\section{Build instructions}\label{sec:build_instr}

\begin{figure}[!ht]
	\centering
	\includegraphics[width=0.9\textwidth]{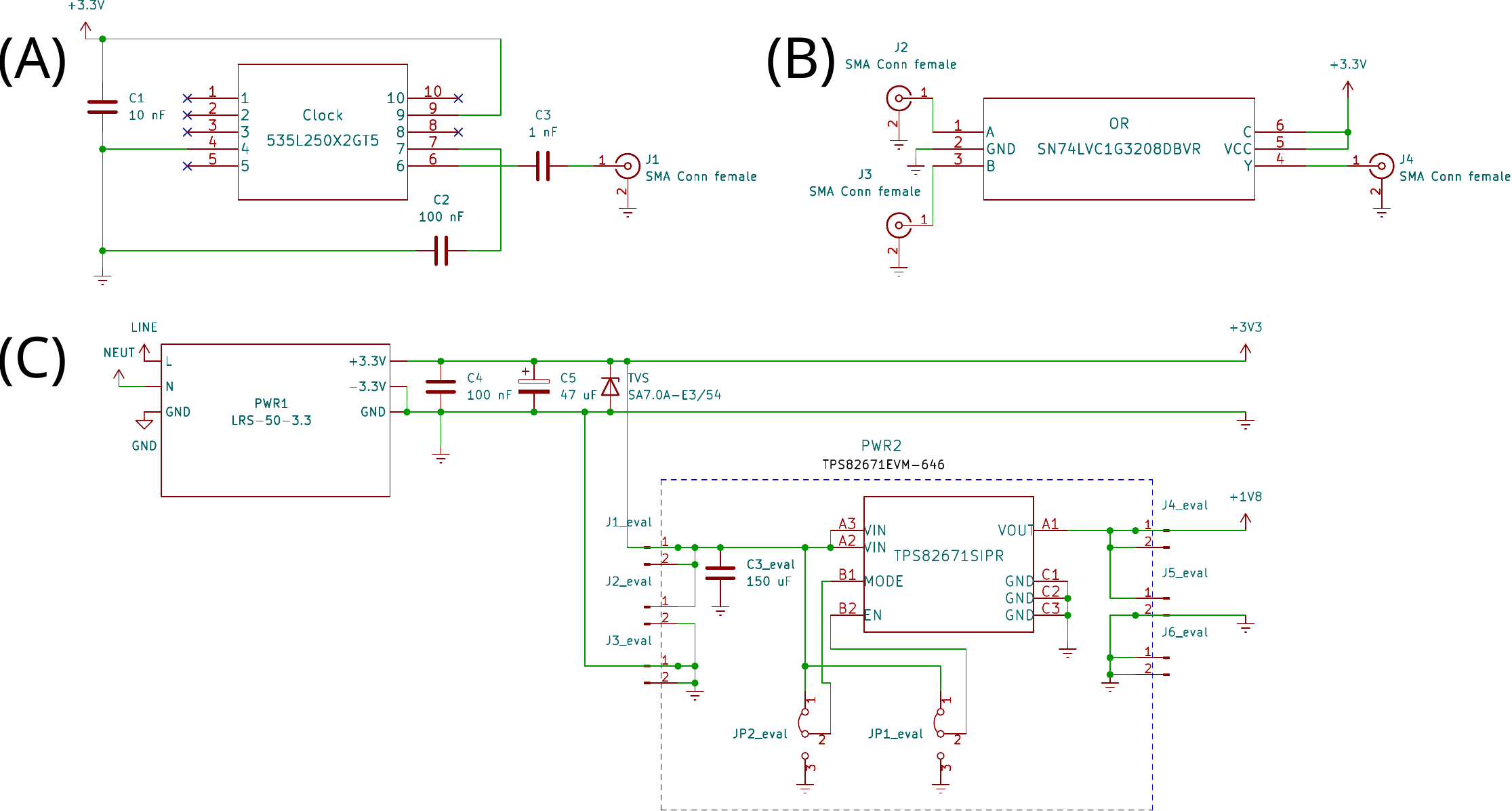}
	\caption{Schematics of (A) the clock, (B) logic OR gate and (C) power supply circuits (the blue dotted line represents the constituent elements of the PWR2 evaluation board).}
	\label{fig:schematicaux}
\end{figure}

In order to power the system we used the circuit depicted in Fig.~\ref{fig:schematicaux}(C), connected using common insulated wires. An AC/DC converter (PWR1) is directly wall powered and generates an output voltage of $+\SI{3.3}{\volt}$, which is filtered by capacitors C4 and C5 (electrolytic) and protected by the TVS Zener diode. For convenience, these three through hole components were soldered and linked on a ProtBoard perforated prototype board. A DC/DC converter (PWR2) is powered from the filtered $+\SI{3.3}{\volt}$ line and produces a $+\SI{1.8}{\volt}$ output when the jumpers JP1 and JP2 are set in the pull-up position. The two voltage levels are necessary to power the DUC (AD9959/PCBZ evaluation board) via the TB1 connector. Additionally, both the OR gate and the clock source circuits adopted in the present application require a $+\SI{3.3}{\volt}$ supply voltage.

The DUC requires an external $\SI{25}{\mega\hertz}$ stable clock reference to operate and reach its peak specifications. We selected a temperature compensated crystal oscillator (Clock) with a frequency stability of $\pm 280$ parts per billion and we assembled the schematic presented in Fig.~\ref{fig:schematicaux}(A) on a ProtBoard, soldering the through hole filtering capacitors (C1, C2, C3) and output SMA jack (J1). An SMA cable (CBsma) delivers the clock signal to the DUC via the J9 connector (REF CLK). In parallel, the W9 jumper had to be set to the REF CLK position. This reference clock was then brought to $\SI{500}{\mega\hertz}$ by employing the internal phase-locked loop-based reference clock multiplier of the DUC, as described in Section~\ref{sec:op_instr}.

The logical OR gate circuit was soldered on the third ProtBoard, following the schematic presented in Fig.~\ref{fig:schematicaux}(B). The OR chip has its two inputs and single output wired to three SMA jacks (J2, J3, J4). The second SMA cable (CBsma) was cut in half and a two position connector (split-off either from the multi-position male header CNm or female receptacle CNf) was soldered on the loose end, with one pin attached to the inner signal line and the other to the outer shielding. The connector type (male or female) should be selected depending on the header type present on the MCU and DUC.

To power the MCU we opted to rely on the USB connector used for programming the board via the cable CBusb, instead of using an external power supply. Alternatively, $\SI{5}{\volt}$ may be supplied via the $\rm{V_{IN}}$ pin. However, for using the USB connection while employing an external power supply, the power provided by the USB cable should be properly isolated, so as to prevent the possibility of power flowing back to the PC. This can be accomplished by cutting apart the $\SI{5}{\volt}$ pads on the bottom side of the MCU board.

\begin{figure}[!ht]
\centering
\includegraphics[width=0.8\textwidth]{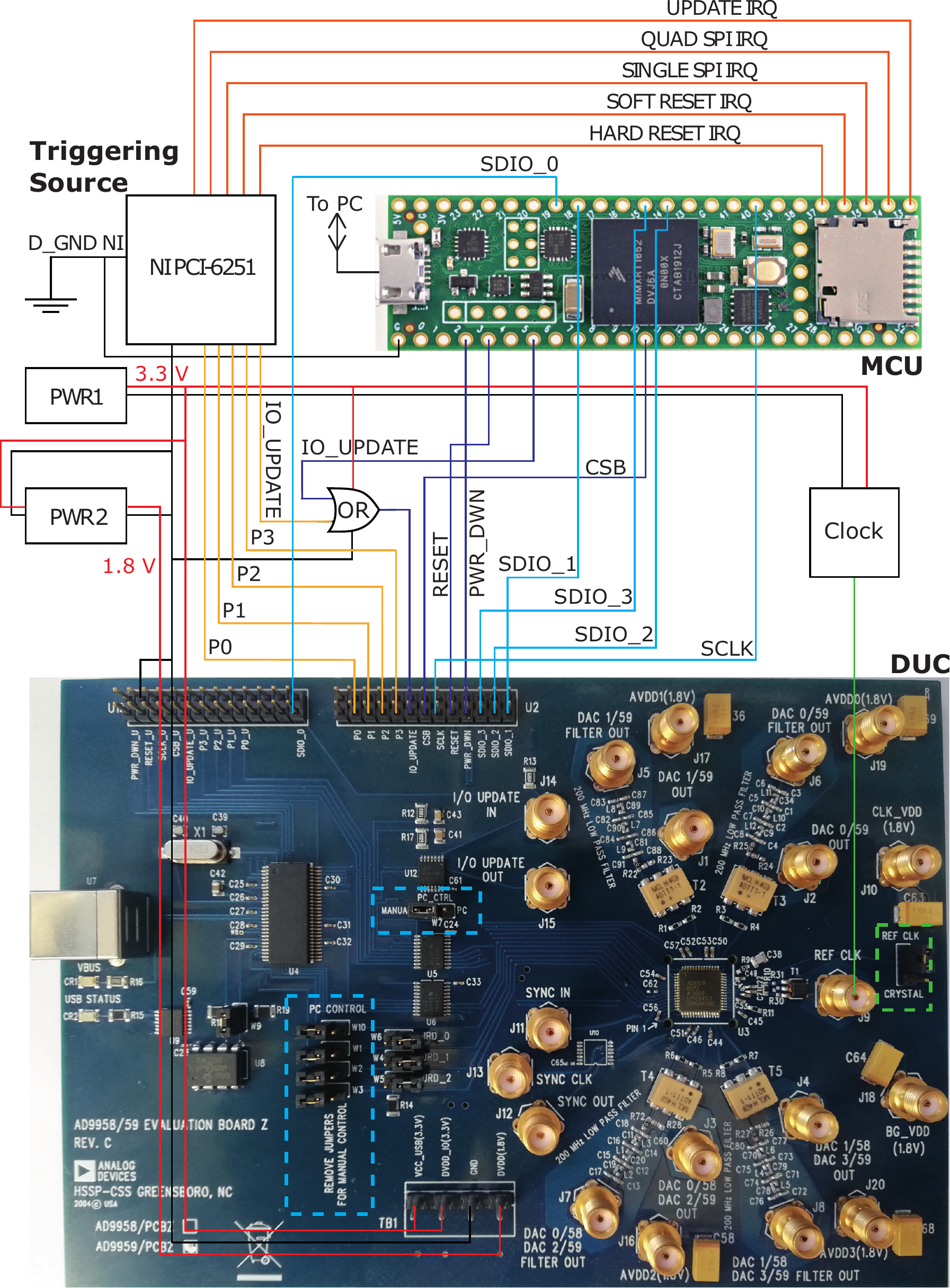}
\caption{System assembly diagram. The positive power supply at $\SI{3.3}{\volt}$ and $\SI{1.8}{\volt}$ is shown in red, with the common ground of all devices shown in black. Digital inputs from the NI board are shown in orange, with dark orange indicating the IRQ lines connected to the MCU and light orange indicating the digital input lines to the DUC. Blue denotes the digital connections between MCU and DUC, with light blue indicating the shielded high-frequency transmission lines employed by the custom SPI developed in this work. The $\SI{25}{\mega\hertz}$ reference clock is shown in green, whereas the jumpers to be disconnected for enabling the manual DUC control and the external clock input are respectively highlighted in blue and green.}
\label{fig:pinconn}
\end{figure}

The MCU can control the DUC by establishing the connections illustrated in Fig.~\ref{fig:pinconn} to the I/O control headers. By default, the AD9959/PCBZ evaluation board is supplied with the USB communication enabled, a setting that disables these headers. To enact the external control of the DUC via the MCU, it is necessary to set the jumper W7 (PC\_CTRL) to manual and remove those on W1, W2, W3 and W10 (highlighted with the blue dotted line in Fig.~\ref{fig:pinconn}). 
Now, the MCU board must be connected to the header row (U2, U13) of the DUC, using, in particular, the PWR\_DWN, RESET, IO\_UPDATE and P0-3 (profile) pins, besides the pins related to the single/quad-wire SPI, i.e. chip select (CSB), serial clock (SCLK), SDIO\_0, SDIO\_1, SDIO\_2 and SDIO\_3. Since in the present application the DUC does not need to send a response back to the MCU board, a MISO (Master Input Slave Output) line was not implemented.

The ground pins next to these connections on the DUC header row must be coupled to the ground references on the MCU for avoiding ground loops. Similarly, the ground of the trigger generator used to control the operation timing must also be connected to the same ground reference. At both ends of each link, one or two position connectors split-off either from the multi-position male header CNm or female receptacle CNf and soldered to an insulated cable can be used to realize mechanically and electrically stable wiring.

The SPI communication between MCU and DUC was implemented in two alternative ways: via the standard single-bit SPI library or using custom functions performing single- or four-bit serial input operations. These functions use direct digital port manipulation for the fast simultaneous control of the designated SDIO and SCLK pins, in order to achieve improved serial transfer times with respect to the standard SPI library available in the Arduino platform. Specifically, all conventional SPI pins belong to the GPIO6 port of the MCU board; conversely, all the remaining pins indicated so far belong to different digital ports. Using these functions, we have verified the possibility of generating a SCLK having a frequency up to $\SI{120}{\mega\hertz}$. However, clocking the SPI chip at this maximum rate could not guarantee a reliable communication, in view of the parasitic elements affecting our system. To mitigate this disturbance, SPI connections were made using shielded cables, grounding the shield at both ends. Adopting a working SCLK rate of $\SI{60}{\mega\hertz}$ by setting a clock divider equal to 2 in the custom SPI functions enabled a reliable communication between the MCU and DUC. The digital pin connections between them are summarized in Table~\ref{tab:pins}.

\begin{table*}[h!]
\centering
\caption{Digital pin connections between MCU and DUC.}
\label{tab:pins}
\setlength{\extrarowheight}{2pt}
\begin{tabular}{lcc}
\hline
\multicolumn{1}{l}{\multirow{2}{*}{\textbf{Function}}}     & \multicolumn{2}{c}{\textbf{Digital pins}}\\
\multicolumn{1}{c}{}  & \textbf{MCU} & \textbf{DUC} \\ \hline \vspace{4pt}
Power Control         & 2   & PWR\_DWN   \\ \vspace{4pt}
Master Reset          & 3   & RESET      \\ \vspace{4pt}
I/O Update            & 5   & IO\_UPDATE \\ \vspace{4pt}
Chip Select           & 10  & CSB        \\ \vspace{4pt}
\begin{tabular}[c]{@{}l@{}}Single/Quad-wire SPI\\ Serial Clock\end{tabular} & 40  & SCLK       \\ \vspace{4pt}
\begin{tabular}[c]{@{}l@{}}Single/Quad-wire SPI\\ SDIO\end{tabular}         & 19  & SDIO\_0    \\ \vspace{4pt}
\begin{tabular}[c]{@{}l@{}}Quad-wire SPI\\ SDIO (second line)\end{tabular}  & 18  & SDIO\_1    \\ \vspace{4pt}
\begin{tabular}[c]{@{}l@{}}Quad-wire SPI\\ SDIO (third  line)\end{tabular}   & 14  & SDIO\_2    \\ \vspace{4pt}
\begin{tabular}[c]{@{}l@{}}Quad-wire SPI\\ SDIO (fourth line)\end{tabular}  & 15  & SDIO\_3    \\
\hline
\end{tabular}
\end{table*}

The last indispensable connections are related to the external timing system which can be any device capable of producing 10 +$\SI{3.3}{\volt}$ TTL signals at rates $\geq\SI{1}{\mega\hertz}$. In our implementation, we used a National Instrument (NI) board (PCI-6251) to generate the interrupt trigger signals related to the commands described in Sect.~\ref{sec:hw_descr:soft_mcu}, in addition to the pulses sent to the IO\_UPDATE and P0-3 pins of the DUC, respectively required to activate the frequency configurations transferred via SPI, and to trigger the generation of the frequency ramps in frequency sweep mode. To meet the $\SI{3.3}{\volt}$ input voltage requirement of both the MCU and DUC boards, the $\SI{5}{\volt}$ output lines of the NI board were connected to a $\SI{5}{\volt}$ to $\SI{3.3}{\volt}$ voltage level translator (\href{https://www.digikey.it/en/products/detail/texas-instruments/SN74LVC4245ADWR/562892}{SN74LVC4245ADWR}). As detailed in Section~\ref{sec:hw_descr:soft_mcu}, we implemented two options to trigger the IO\_UPDATE pin on the DUC: either directly from the NI after the SPI communication is completed or from the MCU if the ``auto update" mode is engaged. To allow both options to work alternatively without hardware modifications, we use the digital logic OR gate circuit whose output goes to the DUC IO\_UPDATE pin, while its inputs are connected to the I/O update pin of the MCU board (pin 5) and to the designated digital output line of the trigger generator.

\section{Operation instructions}\label{sec:op_instr}

To operate the Arduino-based control system we developed an open-source software solution comprising two elements: the Arduino C++ sketch that constitutes the MCU program and the user interface implemented as a Python Jupyter notebook that runs on the PC and communicates with the MCU via USB. This notebook allows to compose and send the frequency configurations to be sequentially generated by the four-channel RF signal generator that we implemented as a demonstration of the MCU control system capabilities. The following procedure should be followed to operate the open-source software controlling the DUC:

\begin{enumerate}
\item Download the complete software package, comprising the Python 3 notebook and the C++ sketch to be loaded on the MCU board, either from the \href{https://dx.doi.org/10.17632/hvwyz5yhh2.1}{Mendeley repository} or from the corresponding \href{https://github.com/lens-biophotonics/open-fast-buffered-4ch-rf-gen}{GitHub repository} \cite{Github} for the latest version.
\item Download and install the Arduino integrated development environment (IDE) (\href{https://www.arduino.cc/en/software}{Arduino Downloads}). 
\item Connect the MCU board to the PC via USB: this will power up the board.
\item Power up the DUC.
\item Open the C++ sketch using the Arduino IDE and, if required, modify the reference clock configuration of the DUC which is preset for an external $\SI{25}{\mega\hertz}$ clock source. In detail, after setting the frequency of the external reference clock (REF\_CLK), set the global c\_ClkMultiplier variable to \textit{false} if clocking the DDS chip directly with a high frequency source. Otherwise, the internal phase-locked loop-based reference clock multiplier will be enabled; in this case, users must also specify the REF\_CLK multiplier factor (c\_PLLmul), that must be between 4 and 20. The internal clock rate of the DDS chip determines the maximum RF frequency that can be generated without incurring in spectrum distortions, which is 40\% of this rate.
\item If desired, activate the sketch debug mode by setting g\_debug to \textit{true} and open the IDE serial monitor. When enabled, the MCU will output to this monitor several debugging messages that may help identifying issues in the supplied configurations or other problems.
\item If desired, activate the optional ``auto update" mode (g\_autoUpdate = \textit{true}), discussed in Section~\ref{sec:hw_descr:soft_mcu}.
\item Use the Arduino IDE to load the C++ sketch on the MCU board.
\item Open the Jupyter notebook user interface.
\item If required, adjust the value of the DUC system clock frequency (SYSCLK) within the ``Setup Constants" cell at the top of the notebook.
\item Verify the number of the serial USB port used to communicate with the MCU board by selecting Tools $>$ Port in the Arduino IDE menu.
\item In order to define the list of single-tone frequencies and/or linear frequency sweeps to be sequentially generated by the four DDS channels upon successive update interrupt events, edit the respective Python lists within  the ``Input to AD9959 DDS" cell. Specifically, whereas a simple floating-point value can be provided for the latter mode of operation, frequency sweeps are configured via Python dictionaries having as keys the start frequency, the chirp parameter (the slope of the frequency ramp, either positive or negative) and the sweep duration. Alternatively, a None keyword may be filled in whenever a particular channel should not be updated upon a specific step of the configuration sequence.
\item If desired, activate the notebook debug mode (debug = True) displaying the Unicode data strings encoded and sent to the board. Moreover, users may optionally disable the serial USB transfer (transfer = False) for validating the input to the MCU board without actually transferring the data strings.
\item Press the notebook's ``Run all" button to transfer the frequency configurations to the MCU board.
\item If the MCU debug mode is active, users may verify in the IDE serial monitor the frequency configurations received via USB and the programming progress of the DUC. Adopt long pulse periods ($\sim$ ms) to allow the PC enough time to receive and display the MCU feedback messages.
\item Send a pulse train to the designated update interrupt pin of the MCU board for sequentially writing the transferred frequency configurations to the SPI buffers of the DUC. Once each setting is uploaded, the DUC waits for an I/O update pulse to reprogram and output the new set of RF signals. This pulse must be synchronized with the pulse train and delayed at least by the duration of the SPI transfer communication. It may originate either from another channel of the trigger source or from the MCU itself if the ``auto update" mode of the C++ sketch has been enabled (item 7).
\item If linear frequency sweeps have been configured, following the I/O update pulse, these need to be externally triggered via transitions of the corresponding DUC profile pin logic state, i.e. from low to high
for rising sweeps and vice versa, as detailed in the Linear Sweep Mode section of the AD9959 data sheet. Such logic state has to last for the entire duration of the frequency sweep (\textit{no-dwell} mode disabled).
\item If required, users may toggle the designated external interrupt pins from low to high in order to trigger the execution of the commands implemented by the ISRs described in Sect.~\ref{sec:hw_descr:soft_mcu}.
\item By repeating items 12-17, additional configuration lists can be defined, transferred to the MCU (appending them to the ones previously sent, if not yet executed) and activated on the DUC.
\end{enumerate} 

\section{Validation and characterization}\label{sec:val}

In order to validate the operation of the Arduino-based control system and of the four-channel RF signal generator that it powers, we first confirmed that the USB and SPI communication steps were error-free and we measured their data throughput rates. Then, we verified that the DUC was correctly programmed for all possible setting combinations and we characterized the RF output spectrum and verified that we could generate the desired combinations of RF single-tone signals and frequency sweeps.

\begin{figure}[!ht]
	\centering
	\includegraphics[width=0.6\textwidth]{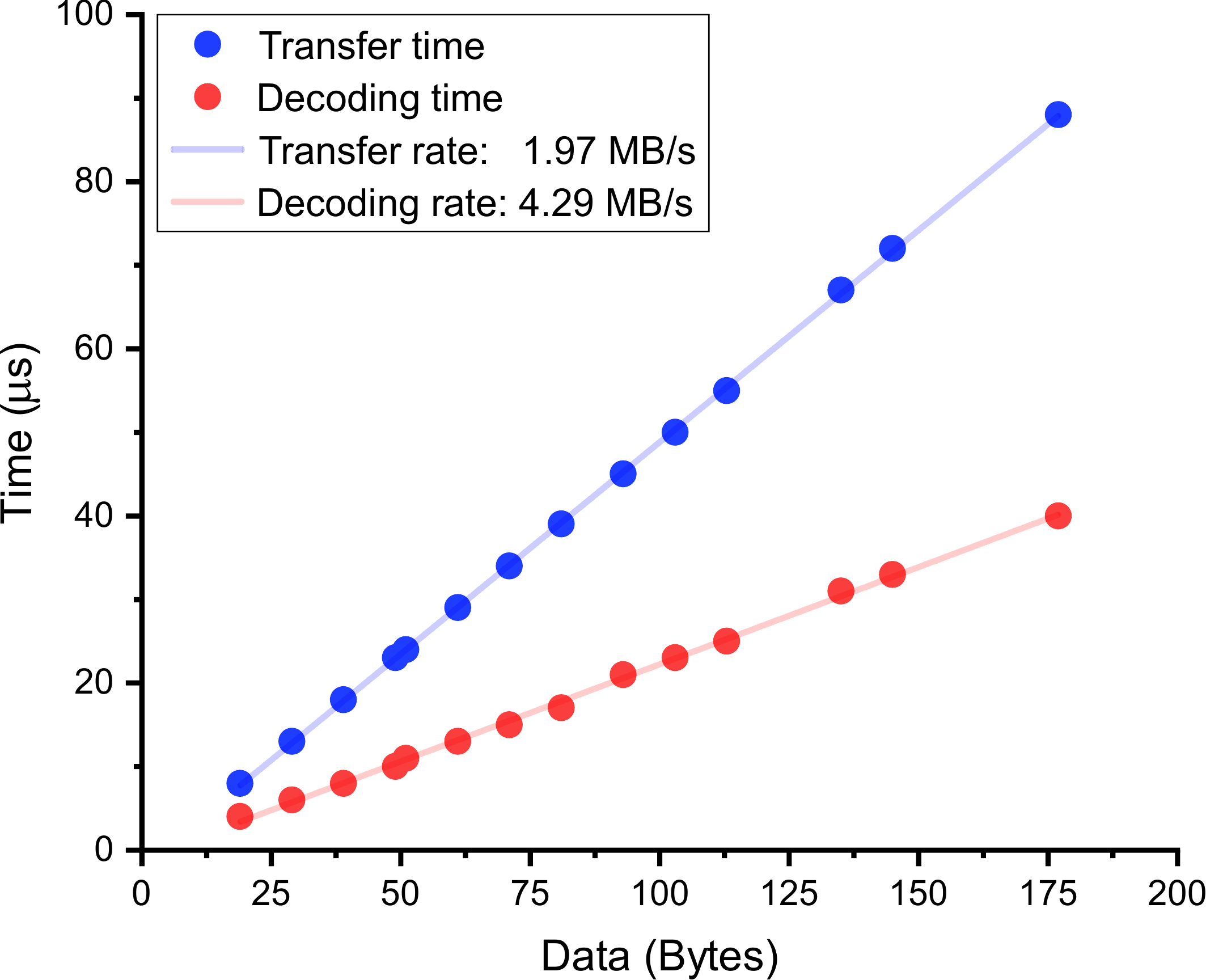}
	\caption{USB communication data rate between PC and MCU board, evaluated by linearly fitting the data reported in Table \ref{tab:usb}.}
	\label{fig:transferUSB}
\end{figure}

We characterized the performance of the serial USB communication between the PC and MCU by evaluating the time required to transfer byte strings related to varying modes of operation of the four DDS channels, along with the time required to parse them and push incoming data into the designated memory buffers (Table~\ref{tab:usb}). The data string byte size increases with the number of channels to be simultaneously activated, especially when operated in frequency sweep mode. We find that the relationship between the amount of transferred data and the incurred time is highly linear, as shown in Fig.~\ref{fig:transferUSB}, and from a linear fit we extract an USB transfer rate of $1.97\pm0.01$~MByte$\si{\per\second}$ and a data decoding rate of $4.29\pm0.01$~MByte$\si{\per\second}$, while the overheads are quite small, respectively $\SI{0.96\pm0.01}{\micro\second}$ and $\SI{0.20\pm0.01}{\micro\second}$.

\begin{table}[!ht]
\small
\centering
\caption{PC-MCU USB communication properties against different channel configurations.}
\label{tab:usb}
\setlength{\extrarowheight}{2pt}
\begin{tabular}{ccrcc}
\hline
\multicolumn{2}{l}{\textbf{Channels operation}} & \multirow{2}{*}{\textbf{Bytes}} & \multirow{2}{*}{\textbf{USB Transfer Time [\textmu s]}} & \multirow{2}{*}{\textbf{Decoding Time [\textmu s]}} \\
Single-tone  & Frequency Sweep  &       &      &     \\ \hline
1            & 0                & 19    & 8    & 4   \\
2            & 0                & 29    & 13   & 6   \\
3            & 0                & 39    & 18   & 8   \\
4            & 0                & 49    & 23   & 10  \\
0            & 1                & 51    & 24   & 11  \\
1            & 1                & 61    & 29   & 13  \\
2            & 1                & 71    & 34   & 15  \\
3            & 1                & 81    & 39   & 17  \\
0            & 2                & 93    & 45   & 21  \\
1            & 2                & 103   & 50   & 23  \\
2            & 2                & 113   & 55   & 25  \\
0            & 3                & 135   & 67   & 31  \\
1            & 3                & 145   & 72   & 33  \\
0            & 4                & 177   & 88   & 40  \\ \hline
\end{tabular}
\end{table}

Next, we similarly assessed the achievable communication rate between the MCU and the DUC against different channel operation configurations, comparing the transfer times allowed by the standard single-wire Arduino SPI library with the ones obtainable with our custom single- or four-wire SPI implementation, which exploits direct port manipulation on the Teensy 4.1, a hardware optimization. A serial clock rate of \SI{60}{\mega\hertz} was consistently adopted for the three SPI communication strategies, in order to establish a proper comparison.

\begin{figure}[!ht]
	\centering
	\includegraphics[width=0.6\textwidth]{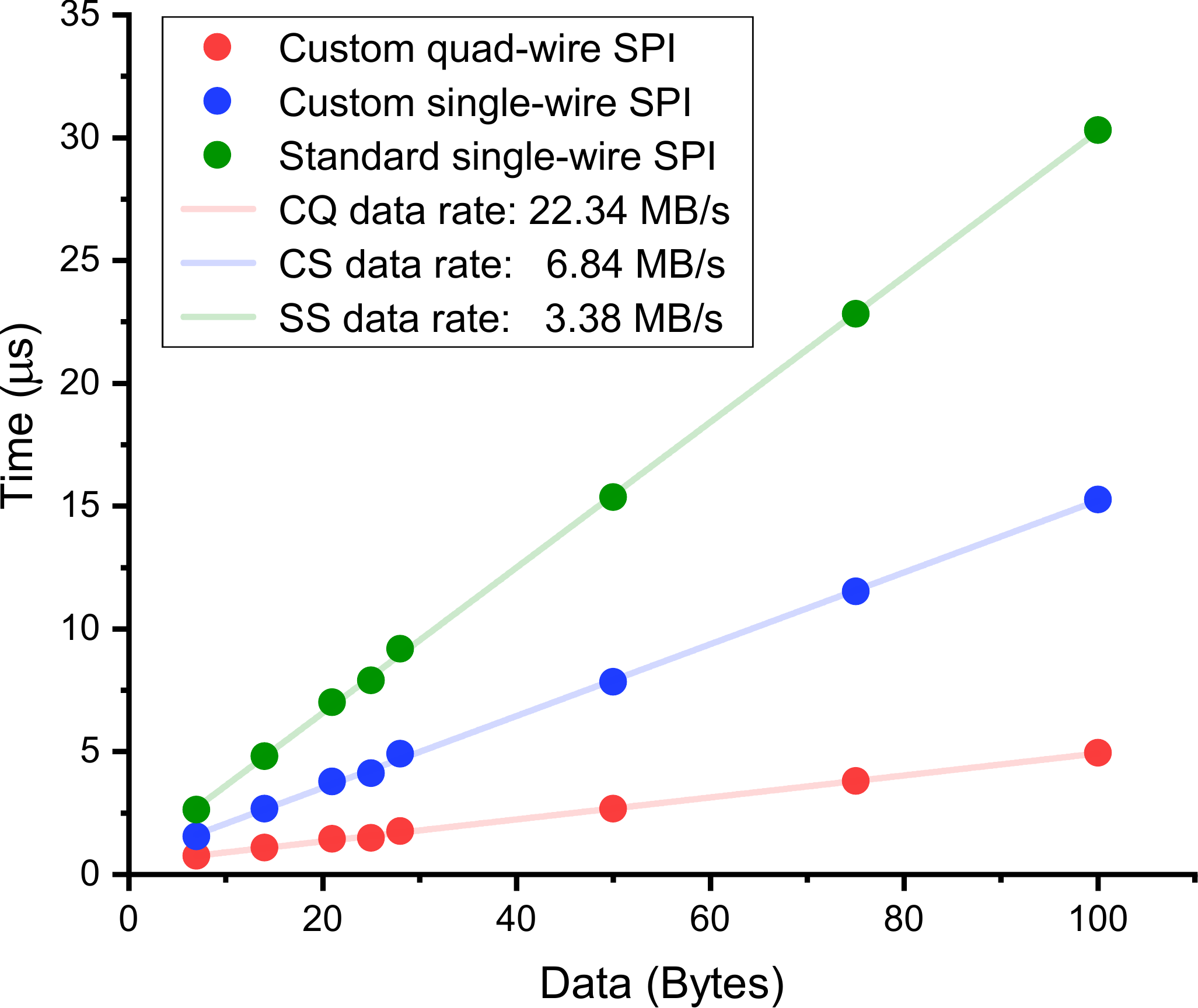}
	\caption{SPI transfer data rate between MCU board and DUC, evaluated by linearly fitting the data reported in Table \ref{tab:spi}, related to the three tested serial operation modes. Measures were performed using a consistent SCLK rate of \SI{60}{\mega\hertz}. A considerable throughput increase is noticeable for the custom quad-wire SPI developed in this work.}
	\label{fig:transferSPI}
\end{figure}

These measures involved a subset of all possible channel programming combinations (i.e., an increasing number of single-tone signals or frequency ramps), and considered also the channel operation mode pre-programming needed when switching at least one of the four DDS channels from single-tone to frequency sweep mode and vice versa, requiring six additional bytes to be transferred in advance to the DUC. The measured transfer times are summarized in Table \ref{tab:spi}, and shown in Fig.~\ref{fig:transferSPI} (channel mode pre-programming excluded). The standard single-wire SPI library provided a data rate of $3.38\pm0.02$~MByte$\si{\per\second}$, while our custom hardware-optimized implementation reached $6.84\pm0.07$~MByte$\si{\per\second}$ in single-wire and $22.3\pm0.3$~MByte$\si{\per\second}$ in quad-wire modes. In all three cases the communication overhead and jitter was very small, respectively of $\SI{192\pm24}{\nano\second}$, $\SI{90\pm10}{\nano\second}$ and $\SI{20\pm2}{\nano\second}$. These results highlight that our custom hardware-optimized SPI implementation delivers a considerable improvement of the DDS input data rate with respect to the standard SPI library, up to $6.6$ times. As reported in Table~\ref{tab:spi}, this in turn leads to an up to 6 times increase in the maximum allowed $ORR$ of the four DDS channels when using quad-wire custom SPI, with a minimum $ORR = \SI{188}{\kilo\hertz}$ in the most data communication intensive case of four pre-programmed frequency sweeps, and a peak $ORR = \SI{1.33}{\mega\hertz}$ when only a single-tone setting is modified. This last result demonstrates that our four-channel RF signal generator reaches the desired performance specification and is competitive with other state-of-the-art solutions presented in Section~\ref{sec:intro}.
 
\begin{table}[!ht]
\small
\caption{MCU-DUC SPI communication times against different output channel configurations with three different transmission modes.}
\label{tab:spi}
\setlength{\extrarowheight}{2pt}
\begin{threeparttable}
\begin{tabular}{lrrrrrrr}
\hline
\multirow{2}{*}{\textbf{Channels operation}} & \multirow{2}{*}{\textbf{Bytes}} & \multicolumn{2}{c}{\textbf{Custom Quad-wire SPI} $\dagger$} & \multicolumn{2}{c}{\textbf{Custom Single-wire SPI} $\dagger$} & \multicolumn{2}{c}{\textbf{Arduino SPI} $\dagger$} \\
            &       & \textbf{Time [\textmu s]} & \textbf{Rate [kHz]} & \textbf{Time [\textmu s]} & \textbf{Rate [kHz]} & \textbf{Time [\textmu s]} & \textbf{Rate [kHz]} \\ \hline
1 Single-Tone                   & 7     & 0.8               & 1333.3          & 1.5                & 649.4            & 2.6             & 380.2         \\
1 Single-Tone $\ddagger$        & 13    & 1.1               & 892.9           & 2.6                & 386.1            & 4.7             & 213.7         \\
2 Single-Tones                  & 14    & 1.1               & 917.4           & 2.7                & 374.5            & 4.8             & 207.9         \\
2 Single-Tones $\ddagger$       & 20    & 1.5               & 684.9           & 3.7                & 268.1            & 6.9             & 145.1         \\
3 Single-Tones                  & 21    & 1.4               & 694.4           & 3.8                & 263.9            & 7.0             & 142.9         \\
3 Single-Tones $\ddagger$       & 27    & 1.8               & 552.5           & 4.8                & 206.6            & 9.1             & 110.4         \\
4 Single-Tones                  & 28    & 1.8               & 565.0           & 4.9                & 203.7            & 9.2             & 108.9         \\
4 Single-Tones $\ddagger$       & 34    & 2.2               & 465.1           & 6.0                & 167.8            & 11.2            & 89.0          \\
1 Frequency Sweep               & 25    & 1.5               & 675.7           & 4.1                & 243.3            & 7.9             & 126.7         \\
1 Frequency Sweep $\ddagger$    & 31    & 1.9               & 526.3           & 5.2                & 193.8            & 10.0            & 100.2         \\
2 Frequency Sweeps              & 50    & 2.7               & 374.5           & 7.8                & 127.7            & 15.4            & 65.1          \\
2 Frequency Sweeps $\ddagger$   & 56    & 3.0               & 328.9           & 8.9                & 112.7            & 17.5            & 57.3          \\
3 Frequency Sweeps              & 75    & 3.8               & 263.2           & 11.5               & 86.7             & 22.8            & 43.8          \\
3 Frequency Sweeps $\ddagger$   & 81    & 4.2               & 239.2           & 12.6               & 79.5             & 24.9            & 40.1          \\
4 Frequency Sweeps              & 100   & 4.9               & 202.4           & 15.3               & 65.5             & 30.3            & 33.0          \\
4 Frequency Sweeps $\ddagger$   & 106   & 5.3               & 188.3           & 16.3               & 61.3             & 32.4            & 30.8          \\ \hline
\end{tabular}
\begin{tablenotes}
\item \footnotesize $\dagger$ SCLK = $\SI{60}{\mega\hertz}$
\item \footnotesize $\ddagger$ channel operation mode pre-programming included
\end{tablenotes}
\end{threeparttable}
\end{table}

\begin{figure}[!ht]
	\centering
	\includegraphics[width=0.8\textwidth]{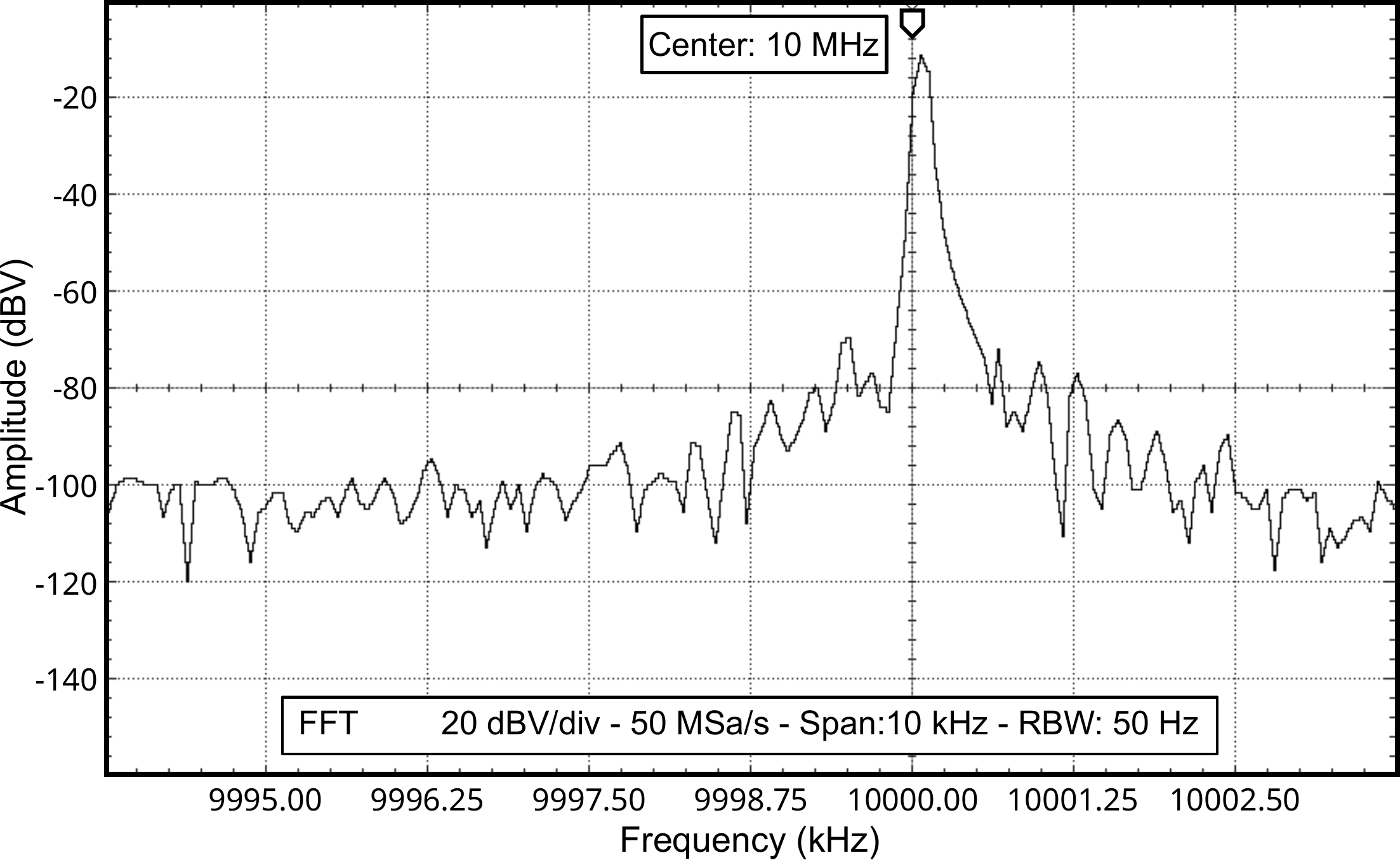}
	\caption{Typical RF spectrum of a single $\SI{10}{\mega\hertz}$ output of the DUC, measured using the Fast Fourier Transform mathematical operation mode of a Rigol MSO5072 oscilloscope.}
	\label{fig:spectrum}
\end{figure}

\begin{figure}[!ht]
	\centering
	\includegraphics[width=0.8\textwidth]{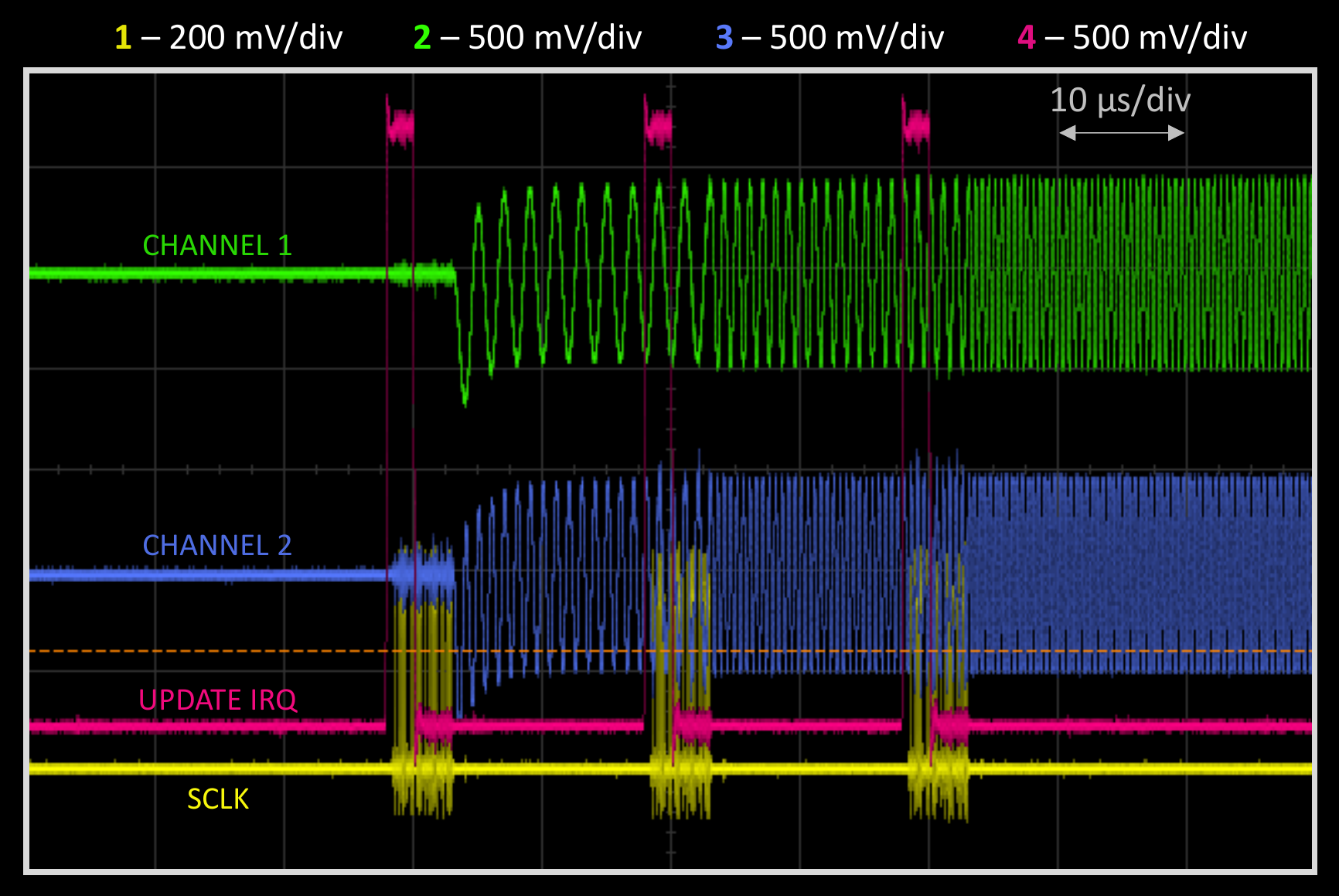}
	\caption{Two-channel signal trace captured with a Keysight Infinivision MSOX2024A oscilloscope, displaying two sequences of three single-tone sinusoids of increasing frequency, activated in response to subsequent rising edges on the designated interrupt pin (UPDATE IRQ, shown in magenta) in ``auto update" mode. The SCLK signal corresponding to the SPI communication between MCU and DUC is shown in yellow.}
	\label{fig:scope}
\end{figure}

Furthermore, we verified the correct activation of the frequency tuning words transferred via SPI, in order to validate the reliability of our custom single- and quad-wire solutions based on fast direct port manipulation. Fig.~\ref{fig:spectrum} shows the spectrum of a $\SI{10}{\mega\hertz}$ RF signal generated by the DUC, as measured using the Fast Fourier Transform mathematical operation mode of a Rigol MSO5072 oscilloscope with a resolution bandwidth of $\SI{50}{\hertz}$. We found the measured spectrum to be consistent with the typical performance characteristics reported in the AD9959 data sheet \cite{AD9959}. The simultaneous refresh of two DDS channels is instead shown in Fig.~\ref{fig:scope}, as captured with a Keysight Infinivision MSOX2024A oscilloscope: three subsequent update interrupt requests detected on the designated digital pin (magenta) trigger the SPI transfer of the data words sequentially stored in the FIFO buffers of the MCU. Following the SPI communication (yellow), an I/O update pulse (not shown) activates the new frequency values which are then generated by the respective DDS output channels after a fixed data latency of a few tens of SYSCLK periods. At the maximum SYSCLK rate of $\SI{500}{\mega\hertz}$, these respectively correspond to $\sim \SI{50}{\nano\second}$ and $\sim \SI{80}{\nano\second}$ for DDS channels operated in single-tone and frequency sweep mode.

\section{Conclusion}

The programmable Arduino-based four-channel RF signal generator that we have developed is an open-source and low cost solution that uses readily available components, like the $\SI{200}{\mega\hertz}$ AD9959/PCBZ \cite{AD9959PCBZ} as the RF DDS source and the Teensy 4.1 \cite{Teensy41} as the microcontroller unit. The total system cost is currently 641.79 \euro, resulting in a notably low cost per channel of 160.45 \euro, which makes it very competitive with respect to both commercial \cite{ModularSystemControls,GraAndAfch,MoglabsQRF,SinaraUrukul,Kasprowicz2022,SpinCoreDDS300,SpinCoreDDS1000,Wieserlabs,WieserlabsDual,MoglabsXRF} and lab-built RF generators \cite{Liang2009,Li2016,Pruttivarasin2015,Perego2018,Prevedelli2019,Donnellan2019,Bertoldi2020,Allcock2021} that have similar specifications and applications. The internal software architecture has been designed to operate as a real-time state machine, allowing it to receive new commands and frequency settings from the PC via USB, store up to millions of them in its internal memory and to almost concurrently reprogram via SPI and activate the DDS outputs with sub-$\si{\micro\second}$ latency and jitter. We have validated the performance of our device, that surpasses all prior MCU-based solutions \cite{ModularSystemControls,GraAndAfch}, and demonstrated that it can generate single-tones or frequency sweeps in an externally or internally triggered arbitrarily programmed sequence. When using our custom quad-wire SPI implementation, that benefits from hardware optimizations, we have achieved high output change rates that depend only on the required amount of reconfiguration data, ranging from a minimum $ORR = \SI{188}{\kilo\hertz}$ for four pre-programmed frequency sweeps and a peak $ORR = \SI{1.33}{\mega\hertz}$ when changing only a single-tone output. These rates were achieved while adopting a SCLK of $\SI{60}{\mega\hertz}$ and may be more than doubled if serial port I/O operations are conducted at the maximum speed of $\SI{200}{\mega\hertz}$ supported by the DUC.
These characteristics make our RF signal generator suitable for a broad range of applications in biophysics, microscopy, quantum, atomic and molecular physics and industrial manufacturing, such as driving acousto-optical devices or controlling the state of processes or matter sensitive to this frequency domain.

By being open-source, state machine based and relying on standard interfaces for communication, the design of our RF generator is easily extensible and customizable for specific applications. Furthermore, the MCU architecture realizes a flexible and complete control system that is adaptable to many other devices. Likewise, the C++ software code presents a low skill barrier to be understood and modified using the Arduino IDE. Perspective improvements of our RF source would be the addition of amplitude and phase control which requires only additional software development, the enhancement of the SPI clock rate to further increase the $ORR$ and the augmentation of command memory by 16~MByte by soldering two extra RAM chips. The functionality of the Jupyter notebook can be expanded to provide an iterative MCU communication method to supply configuration sequences whose length surpasses the MCU memory, to receive status information from the MCU and to provide IRQ functionality via USB, lowering the amount of required TTL timing lines. Finally, it should be possible to design a more general purpose MCU software code, lacking RF generator-specific functionality, that could ease the adaptation to another DUC type interfaced via SPI. Such future improvements and bug corrections will be made available in our \href{https://github.com/lens-biophotonics/open-fast-buffered-4ch-rf-gen}{GitHub repository} \cite{Github}.\\


\noindent
\textbf{CRediT author statement}\\
\noindent
\textbf{Michele Sorelli}: Software, Investigation, Validation, Visualization, Writing - Original Draft, Writing- Reviewing and Editing. \textbf{Marco Marchetti}: Software, Investigation, Validation, Writing - Original Draft. \textbf{Pietro Ricci}: Investigation, Validation, Visualization. \textbf{Domenico Alfieri}: Conceptualization, Funding acquisition, Project administration, Writing- Reviewing and Editing. \textbf{Vladislav Gavryusev}: Conceptualization, Methodology, Supervision, Software, Investigation, Visualization, Writing - Original Draft, Writing- Reviewing and Editing. \textbf{Francesco Saverio Pavone}: Funding acquisition, Project administration, Writing- Reviewing and Editing.\\

\noindent
\textbf{Declaration of Competing Interest}\\
\noindent
The authors declare that they have no known competing financial interests or personal relationships that could have appeared to influence the work reported in this paper.\\

\noindent
\textbf{Acknowledgements}\\
\noindent
The authors would like to acknowledge the help of Dr. Giuseppe Sancataldo with the development of the ATTRACT project \cite{Ricci2022}.\\
\noindent
This project has received funding from the ATTRACT project funded by the EC under Grant Agreement 777222 and from the H2020 EXCELLENT SCIENCE - European Research Council (ERC) under grant agreement ID n. 692943 BrainBIT. This project has also received funding from the European Union’s Horizon 2020 Framework Programme for Research and Innovation under the Grant Agreement No. 871124 (Laserlab-Europe), and was supported by the EBRAINS research infrastructure, funded from the European Union’s Horizon 2020 Framework Programme for Research and Innovation under the Specific Grant Agreement No. 945539 (Human Brain Project SGA3). This research has also been supported by the Italian Ministry for Education, University, and Research in the framework of the Advance Lightsheet Microscopy Italian Mode of Euro-Bioimaging ERIC. Vladislav Gavryusev has been funded by a Marie Skłodowska-Curie Fellowship (MSCA-IF-EF-ST ``MesoBrainMicr" grant agreement No. 793849).\\

\bibliographystyle{plainurl}
\noindent\bibliography{manuscript_HardwareX} 

\end{document}